# Fast Generation of Representative Synthetic Dataset with Salsa to Train ATR Models with Electromagnetic Couplings Data-Augmentation

Benjamin Camus
Scalian DS
2 rue Antoine Becquerel
35700 Rennes, France
benjamin.camus@scalian.com

Julien Houssay
Scalian DS
2 rue Antoine Becquerel
35700 Rennes, France
julien.houssay@scalian.com

Corentin Le Barbu
Scalian DS
2 rue Antoine Becquerel
35700 Rennes, France
corentin.lebarbu@scalian.com

Eric Monteux
Scalian DS
2 rue Antoine Becquerel
35700 Rennes, France
eric.monteux@scalian.com

Cédric Saleun
DGA Maîtrise de l'Information
BP 7 35998 Rennes CEDEX 9, France
cedric.saleun@intradef.gouv.fr

Jean-Christophe Louvigné
DGA Maîtrise de l'Information
BP 7 35998 Rennes CEDEX 9, France
jean-christophe.louvigne@intradef.gouv.fr

***Abstract*—This work focuses on training Automatic Target Recognition (ATR) models using simulated Synthetic Aperture Radar (SAR) images to circumvent the lack of real measurements. To obtain robust and versatile ATR models, simulation needs to generate massive datasets that encompass all the variability found in real measurements. Thus, we need a simulator that finds a good tradeoff between execution speed, computational resource consumption, and physical representativeness. In this work, we demonstrate that the Salsa simulator addresses this issue. We ran computing performance tests to show that Salsa can generate 21,600 synthetic images in less than 10 minutes using a single Nvidia GeForce RTX 4090 GPU. Using our ADASCA Deep Learning approach, we demonstrate that these data are sufficiently representative to train ATR models and reach state-of-the-art results on the MSTAR public dataset with an accuracy of 86 %. To illustrate how Salsa unlocks new possibilities to train ATR models, we use the simulator to conduct a study on Electromagnetic (EM) couplings between the targets and their immediate environment. We demonstrate that, if not accounted for in the training dataset, the variability of the EM couplings induced by the variability of the ground surfaces can significantly degrade the performance of ATR models, with an accuracy decrease of more than 4 %. We also show that Salsa can generate in a timely manner (i.e., in less than 4 hours using the same GPU as previously) a massive dataset of 648,000 images with a large variety of couplings to make the ATR models robust to EM coupling variations. Our ATR models can then achieve an accuracy of 87 % on the MSTAR dataset.**

## I. INTRODUCTION

Synthetic Aperture Radar (SAR) is an active imaging technique producing images with coherent microwaves. It requires a moving antenna (usually installed on a plane or a satellite) that transmits and receives a sequence of electromagnetic (EM) pulses. Images are then formed thanks to signal matched filter processing by exploiting the Doppler effects induced by the motion of the antenna and the frequency ramp of the transmitted signal. This process is called image focusing.

The advantages of SAR compared to traditional optical imagery are numerous. Notably, images can be produced day or night, regardless of weather conditions (e.g. depending on the wavelength used, the signal can penetrate clouds and vegetation to image the area underneath) with a dominant backscattering response from manmade metallic structures. It is then particularly well-suited for the remote sensing of targets in a Defense context.

Automatic Target Recognition (ATR) on SAR images is a long-standing problem that consists of automatically classifying objects of interest. Numerous studies have demonstrated that Deep Learning is able to tackle this challenge [1][2]. However, Deep Learning requires thousands of labelled data to train ATR models. Acquiring such a diversified and complex dataset may be very expensive, or even infeasible, especially in Defense applications. Indeed, because targets signatures vary significantly with the observation geometry, the training datasets must cover a wide angular range, with a sufficiently fine angular resolution—particularly in azimuth. Moreover, the signatures depend on the radar sensor (e.g. carrier frequency, bandwidth, thermal noise), and on the environment of the target, which may also vary greatly (e.g. electromagnetic coupling, shadowing effects). Finally, the targets themselves may be highly variable because they are equipped with articulated parts (e.g. gun turret, doors) and removable components (e.g. fuel tanks).

To build robust and efficient ATR models, the training datasets must encompass all these variation factors for all targets, including potential confusing objects, and full range systematically explore their extent as much as possible, while providing reliable ground truth information to label images. Acquiring enough real measurements to capture this combinatorial explosion is infeasible in practice. This makes SAR simulators the only viable alternative to build synthetic datasets. With simulation, we have full control over the 3D models of the targets and their virtual environment. It is then possible to produce variants at will, with different electromagnetic (EM) materials, and to run parametric production to automatically consider a large amount of observation geometries and associate ground truth information. Simulation can then capture the diversity observed in real operational scenarios.

However, to unlock the full potential of simulation to generative massive training datasets, we need to find a good trade-off between execution speed, computing resources usage, and physical representativeness. Simulations based on "exact" EM models are notoriously very computing-intensive and requires significant execution times. Such approaches can only produce small datasets, capturing only a limited number of real-world variations. Besides, the representativeness of these simulations is often limited by the quality of the input scene, as high-quality 3D models that precisely match the ground truth are rarely available. Thus, simplifying assumptions are required. However, fast simulators based on simplistic models neglect complex EM effects and produce datasets that are not representative of real measurements. In both cases, ATR models trained on such synthetic datasets will have poor generalization ability on real measurements due to the well-known Dataset Shift problem, which occurs when the training and test distributions are too dissimilar [3]. As shown on Fig. 1, the distribution generated by a slow but representative simulator (with representative input scenes) fails to encompass all the real measurements distributions, whereas the distribution generated by a fast but oversimplified simulators don't overlap the target distribution.

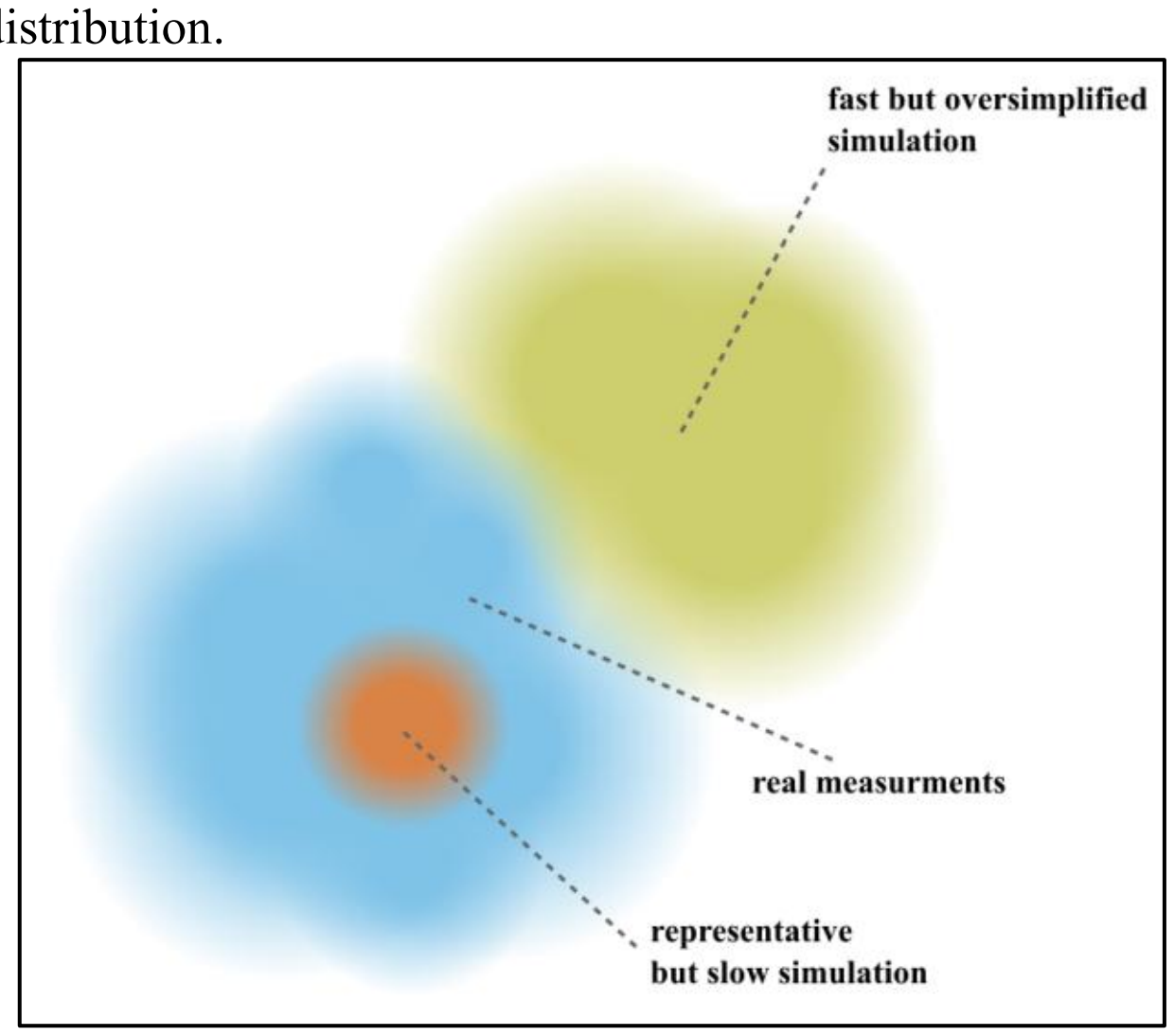

Fig. 1. Simplified view of the dataset shift problem for ATR trained with synthetic data.

In this paper we demonstrate that the Salsa SAR simulator [4] answers this problem. Salsa adopts a Shooting and Bouncing Rays (SBR) technique with a combination of Geometrical and Physical Optics (GO and PO) to compute the EM signature of the targets. The computations are massively parallelized on GPU using NVIDIA OptiX and CUDA. It also adopts a so-called "fast image" rendering strategy in which the EM signatures are directly evaluated at the middle of the integration sector and rasterized on an image grid. This approach is significantly more computationally efficient than the one used by most of the other SAR simulators that consist of computing a succession of radar pulses all along the integration sector and focusing the obtained signal to get the final image. Here, we conduct performance tests to show that Salsa can generate 21,600 synthetic images in less than 10 minutes using a single Nvidia GeForce RTX 4090 GPU. We also demonstrate that these data are sufficiently representative to train ATR models using our ADASCA approach [5] and reach state-of-the-art results on the MSTAR public dataset.

To illustrate how Salsa unlocks new possibilities for training ATR models, we conduct a study on EM couplings with the simulator. Coupling effects correspond to EM interactions between an object and its immediate environment (i.e. "multiple bounces" of the EM waves between the target and the ground). They significantly impact the EM signature of the target. The problem is that they are highly variable because they are very dependent on the environmental configuration. This means that, depending on the ground features (e.g. rugosity, composition), the target signature will change significantly in the SAR image, which greatly complicates the ATR task. Thanks to the computing performance of Salsa, we generated (using the same single GPU as previously) 30 variants of the target signatures for each of our 21,600 images by considering different ground surfaces in our 3D scene in less than 4 hours. By comparing the performance obtained on the MSTAR data with ATR models trained separately with these different variant datasets, we first demonstrate and quantify the impact of ground rugosity on the ATR performance. Then, we show that all these variants can be used simultaneously to train ATR models to make them more robust. Thus, we show that Salsa unlocks both new experiments and new technics for improving ATR.

The rest of the paper is organized as follows. Section II reviews the related works for the ATR approaches in the literature, the existing simulation approaches and the publicly available SAR ATR datasets. Section III introduces the tools used in our approach, i.e. the Salsa simulator and the ADASCA Deep Learning algorithm. Section IV discusses the production of our different datasets. Finally, Section V details our experiments and results regarding both the computing performance of Salsa and the EM couplings study.

## II. Related Works

### A. ATR Datasets

The MSTAR (Moving and Stationary Target Acquisition and Recognition) public dataset [6] comprises measured SAR images of fifteen different targets acquired at different depression and azimuth angles by an airborne radar. Following the standard experiments evaluation plan, the 3671 images collected at a depression angle of 17° constitute the training set whereas the 3203 images with a depression angle of 15° serve to test the models. These data concern ten classes of vehicles (2S1, BMP2, BDRM2, BTR60, BTR70, D7, T62, T72, ZIL131 and ZSU23-4), measured almost at each azimuth degree from 0° to 360°. Three variants of vehicles are available for two classes: BMP2 and T72. Vehicles equipment and configuration (e.g. side skirts) slightly differ across variants.

The SAMPLE (Synthetic and Measured Paired Labeled Experiment) public dataset [7] comprises pairs of real SAR measurements and simulated images. This dataset is smaller than MSTAR with only 806 paired synthetic and real measurements for training (at depression angles of 14°, 15° and 16°) and 539 pairs for testing (at depression angles of 17°). Like MSTAR, SAMPLE comprises ten target classes labelled 2S1, BMP2, BTR70, M1, M2, M35, M60, M548, T72, and ZSU23-4. SAMPLE data does not provide any variant for any class. The SAMPLE dataset suffers from several drawbacks. First, the azimuth angles of the images range only from 10° to 80° for both training and test datasets. It particularly excludes the cardinal directions that may be the more challenging angles. Secondly, this angular sector may not be representative of the challenges encountered when classifying images at a full 360° extent. In addition, the SAMPLE authors have made considerable efforts to make the synthetic data as close as possible to real measurements, using detailed ground truth information. This very favorable scenario is unlikely to occur in an actual operational context where the measured vehicles may differ significantly from the 3D models used during training.

### B. SAR simulation

SAR simulators can be broadly divided into two main categories: rigorous full-wave EM simulators (e.g., MoM, FEM, FDTD), which offer high physical fidelity at the cost of significant computational resources, and high-frequency fast simulators based on asymptotic methods such as GO, PO and related hybrid ray-based or diffraction-based approaches. Only the latter category enables the generation of large-scale SAR datasets suitable for ATR development within practical computation times, while still capturing essential scattering phenomena, polarization effects, and coherent phase information.

Within this category, several SAR simulators are based on the ray-tracing techniques such as RaySAR [8], XPatch [9] [10], and SARCASTIC [11]. They allow to generate SAR raw data composed by a succession of radar pulses that can be post-processed using off-the-shelf focalization algorithms to form SAR images. Amongst these simulators, SARCASTIC is the fastest, with near real-time simulation performance using 2 V100 Nvidia GPUs (i.e. a few seconds per image to compute the radar pulses, plus the time required by the focusing step) [11].

Unlike these ray tracing approaches, MOCEM [12][13] developed by Scalian DS for the DGA (French MoD) only searches for the canonical radar effects that dominate the target signature (namely exact or approximate diffuse, plate, dihedral, trihedral) as well as their mirror counterparts. Their contributions are computed from analytical equations, using PO for the last facet and GO for the other ones. The effects detection is based on an original algorithm that performs an exact geometrical search with facets projection. Contrary to a ray tracing approach, this algorithm ensures that all the small elements of the 3D model are considered, and thus that no important EM contributions are ignored. Once detected, the scatterers are filtered according to geometric criteria (e.g., facets orthogonality) to speed up the calculations and to consider only the canonical effects that dominate the target signature and whose contribution can be computed thanks to an analytical model. In the "fast image" mode, the detected effects are rasterized in a focusing grid to directly form the image. This eliminates the need to simulate all the radar pulses (only the position of the antenna at the middle of the integration sector is considered) and the focusing step. This significantly speeds up simulations, but neglects effects inherently linked to SAR azimuthal/angular integration and to the focusing algorithm (e.g., phase evolution, gradual shadowing, defocusing artifacts). Due to its geometrical search algorithm, MOCEM runs only on CPU hardware, which makes it slower than GPU-based ray tracing approaches for a single run (depending on the number of facets, MOCEM takes a few minutes to generate a single image) but allows for massive parallelization of different runs on several CPU core (e.g. 32 runs can be typically parallelized on a single Intel Core i9).

### C. ATR approaches

Ødegaard et al. got mixed results (65 % of accuracy) when training an off-the-shelf deep-learning algorithm directly on simulated SAR data [14]. Using a transfer learning strategy, Malmgren Hansen et al. pretrained an ATR classifier on a large amount of synthetic data before training the model on a smaller set of measured images [15]. The limitation of this approach is that it still requires measured data for training, and in many cases, measured images of the target of interest may not be available.

Several studies focus on learning an optimal transport function to refine synthetic data by adding features peculiar to measured images. The goal is to transport the synthetic distribution to the measured one to address the dataset-shift issue. The refined synthetic dataset can then be used to train a conventional classifier. Cha et al. trained a residual network to refine synthetic data for ATR [16]. However, their classifier only achieved an accuracy of 55 % at test time.

Lewis et al. [17] and Camus et al. [18] trained a GAN to refine synthetic SAR images, with promising results of almost 95 %. However, all these refinement approaches require real measurements that may be impossible to obtain. Moreover, Camus et al. demonstrated that GANs cannot refine previously unseen classes [18].

Inkawhich et al. trained ATR classifiers on the public synthetic SAMPLE dataset by combining several algorithms from the literature designed to improve the generalization of deep-learning models [19]. They evaluated all their models on the SAMPLE measured images. With several combinations of these techniques the authors achieved an accuracy of almost 95 % on the measured images. However, the authors used the SAMPLE dataset that contains several flaws, as stated in the previous section. Camus et al. [5] have demonstrated that the results drop to 66 % when considering a slightly less favorable scenario. Therefore, these approaches may not be relevant in actual operational scenario.

Delhommé et al. [20] used physics-based data augmentation techniques based on Attributed Scattering Centres (ASC) parameters. By training their ATR models only on synthetic data generated by the MOCEM simulator, they reached an ATR accuracy of almost 71 % on MSTAR.

## III. Our Approach

### A. Salsa

Salsa was designed to rapidly compute the EM signature of meshed objects derived from 3D models, with the objective of updating this signature dynamically during a time-domain simulation such as producing radar raw IQ signals. The long-term goal is to enable the recalculation of environment–target couplings at each radar pulse for dynamically moving scenes (like a boat on the sea surface). These objectives require a careful balance between physical representativeness and computational performance.

For this reason, Salsa adopts the Shooting and Bouncing Rays (SBR) technique, implemented in a dual architecture: a CPU-based version using Intel Embree for the prototyping phase and a GPU-accelerated version using NVIDIA OptiX and CUDA for the production phase.

The EM model relies on the classical assumptions of GO to determine ray trajectories, while PO is applied to the final interaction facet to compute the complex reflected field, considering the properties of dielectric materials and the polarization effects.

Through these design and implementation choices, Salsa provides a fast and physically coherent framework to model complex manufactured targets producing multiple bounces. It is fully compatible with both monostatic and bistatic radar configurations. The ray-launch density plays a critical role in controlling the trade-off between accuracy and computing speed. A distinctive feature of Salsa is that it allows the generation of multiple types of radar products and offers a complementary and versatile approach to SAR image formation. For a given acquisition geometry, Salsa can compute the complete complex scene response, enabling holographic reconstruction through angular and frequency synthesis. It can also generate range profiles suitable for IQ data generation, which can subsequently be focused using algorithms such as Back-Projection. Both approaches can be used in bistatic configurations.

Alternatively, in monostatic configurations, Salsa can produce "fast images" similar to MOCEM. With this mode, Salsa rasterizes the backscattered rays into a focusing grid to form a so-called "source image". The source image is an ideal SAR image without thermal noise, and with an impulse response that corresponds to a Dirac delta function. The source image can then be converted into a proper monostatic radar image by applying the transfer function of the sensor (see Fig. 2), which is based on the image quality to reproduce (i.e. resolution, side-lobes, thermal noise). Similarly to MOCEM, this mode is significantly faster than the traditional raw IQ data generation scheme because it requires only one raytracing step to produce an image instead of one raytracing for each radar pulse (a typical acquisition being composed of hundred, of even thousands of pulses), and because it does not need to apply a focusing step. However, it still neglects effects linked to SAR azimuthal/angular integration. In this study, we use this fast image mode to fully take advantage of the computing efficiency of Salsa.

Surface roughness is introduced by using a meshed rough surface for the ground. The main purpose is to sufficiently perturb the coupling effects to reduce their overly coherent and directive behavior that occurs with a smooth coupling surface. This introduces a more realistic diversity of target–ground interactions, better representing the variability of a rough terrain. Salsa provides a dedicated module called NACHOS (Numerical Algorithm for Computing Heightmaps Of rough Surfaces) to generate such rough meshes.

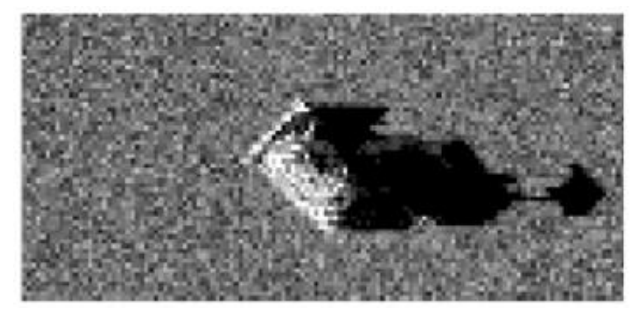
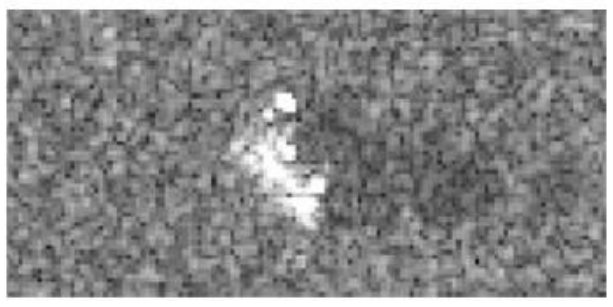

Fig. 2. Source image (left) vs. radar image (right) of the D7 target

### B. ADASCA

ADASCA was specifically designed to train ATR models on synthetic data [5][21]. It uses Adversarial Training (AT) [22] to improve the generalization of the ATR models. This AT strategy consists of altering at runtime the training images with small perturbations to lower the model predictions as much as possible. To this end, we use the Fast Gradient Sign Method (FGSM) [23] to compute the optimal noise to apply to the images.

We combine AT with an intensive physics-based domain randomization strategy [24]. It consists of randomizing simulated environment parameters to introduce as many variations in the synthetic data as possible. For each training epoch, we create a variant of our synthetic dataset by randomly determining separately for each image (see Fig. 3): the range and cross-range resolution, the level and distribution of the background clutter, the thermal noise of the sensor, and the target position in the images.

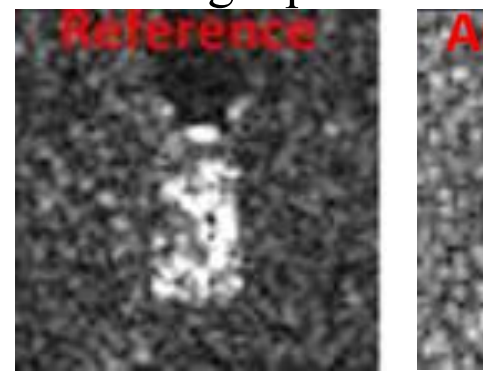

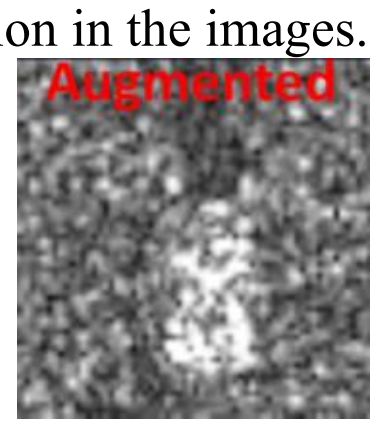

Fig. 3. Example of domain randomization.

ADASCA applies these data augmentations directly on the source images generated by Salsa. It then generates and applies then the sensor function to convert the source images to radar images according to the desired image quality. ADASCA works on oversampled source images to consider the sub-pixel position of the scatterers, improving significantly the representativeness of the data for a low computational cost. The images are downsampled after applying the sensor function to match the expected pixel size. ADASCA is implemented in Tensorflow (TF) to run in parallel on a GPU and to perform all these data augmentations on batches of images directly at runtime during the classifier training.

We use the bagging method [25] to average the prediction of five independent models. We also perform Test-Time Data Augmentation (TTDA) [26]: at inference-time, we create 20 variants of each test image by applying random circular shifts in the range of [-5, 5] pixels range in the x and y dimensions. Then, we average the model predictions to classify the image.

We use a DenseNet121 architecture [27], an SGD (Stochastic Gradient Descent) optimizer with Nesterov momentum [28], a weight decay of $10^{-4}$, and a batch size of 512. The learning rate and the momentum vary during the training according to a "1cycle" policy [29]. We train each model for 300 epochs.

## IV. Dataset production

To generate our datasets with Salsa, we consider the 10 targets of the MSTAR dataset. Unlike previous studies in the literature, we do not fine-tune our simulations (i.e. 3D models and EM materials) to closely match the ground truth of the test data. This brings us closer to an operational scenario.

We take off-the-shelf 3D models available on the Internet. We use one 3D model per class (see Fig. 4). We assign generic EM materials (i.e. with well-defined reflectivity and dielectric constant) with the different facets of the model. We consider the transfer function of the MSTAR sensor (i.e. with similar range/cross-range sampling and resolution, thermal noise level, and Taylor window function). With Salsa, we run parametric productions for the 16°, 17° and 18° depression angles. For each depression, we generate images at every 0.5° azimuth for the full 360° range. Thus, our training dataset comprises a total of 21,600 images.

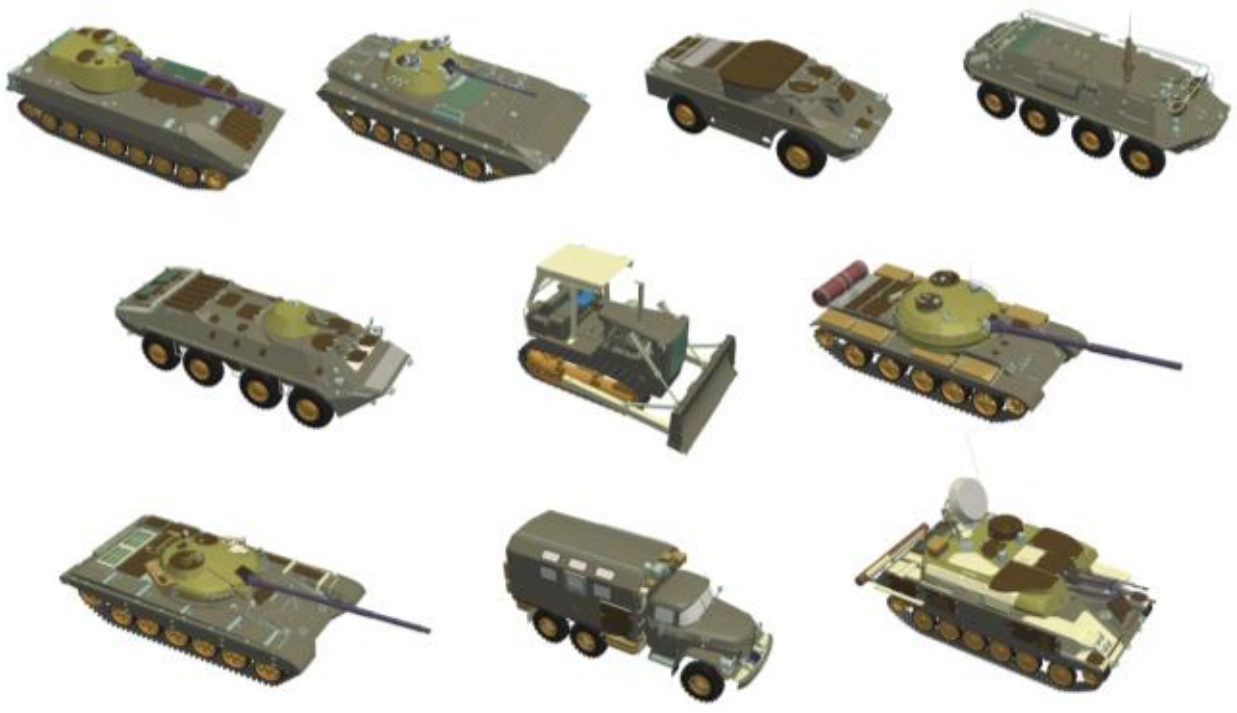

Fig. 4. 3D models of the MSTAR targets used in our simulations. From top left to bottom right : 2S1, BMP2, BRDM2, BTR60, BTR70, D7, T62, T72, ZIL131, ZSU23-4.

We tested our ATR models on MSTAR data acquired at 15° depression angles (including the T72 and BMP2 variants). Thus, we do not consider the exact same azimuth and incidence angles for training and test.

To simulate variations of the EM couplings, we used the NACHOS module to generate different meshed rough surfaces for the ground (see Fig. 5). A shown in TABLE I. , we considered 9 different Gaussian rough surfaces that vary according to the height standard deviation ( $\sigma h$ ), and correlation length ($cl$). We considered typical values for these variables [32][33]. We used 3 variants of materials to characterize the EM reflectivity of these surfaces (see TABLE II. ). These values are representative of increasing moisture conditions [34]. Gaussian rough surfaces constitute a first-order, simplified model that is useful for the present study. This approach can be further extended to more realistic and structured surface representations, including multi-scale models or stochastic fields such as Perlin noise.

TABLE I. ROUGH GROUND SURFACE GENERATED

| ID | σh (cm) | cl (cm) | Slope (%) | Description |
|---|---|---|---|---|
| G1 | 0.4 | 12 | 4.7 | smooth, e.g. tarmac |
| G2 | 0.8 | 12 | 9.4 | Short grass |
| G3 | 1.2 | 12 | 14.1 | Moderately rough bare soil |
| G4 | 1. | 30 | 4.7 | Gently undulating |
| G5 | 2. | 30 | 9.4 | Rolling terrain |
| G6 | 3. | 30 | 14.1 | Moderately rugged |
| G7 | 4.5 | 30 | 21.2 | Strongly undulating |
| G8 | 6.0 | 30 | 28.3 | Highly rugged |
| G9 | 8.0 | 30 | 37.7 | Extreme roughness |

TABLE II. EM MATERIALS USED FOR THE GROUND

| ID | $\epsilon$ | $\mu$ | Description |
|---|---|---|---|
| M1 | $3 - 0.05i$ | 1 | Dry ground |
| M2 | $6 - 0.5i$ | 1 | Moderately wet soil |
| M3 | $15 - 1.5i$ | 1 | Strongly wet soil |

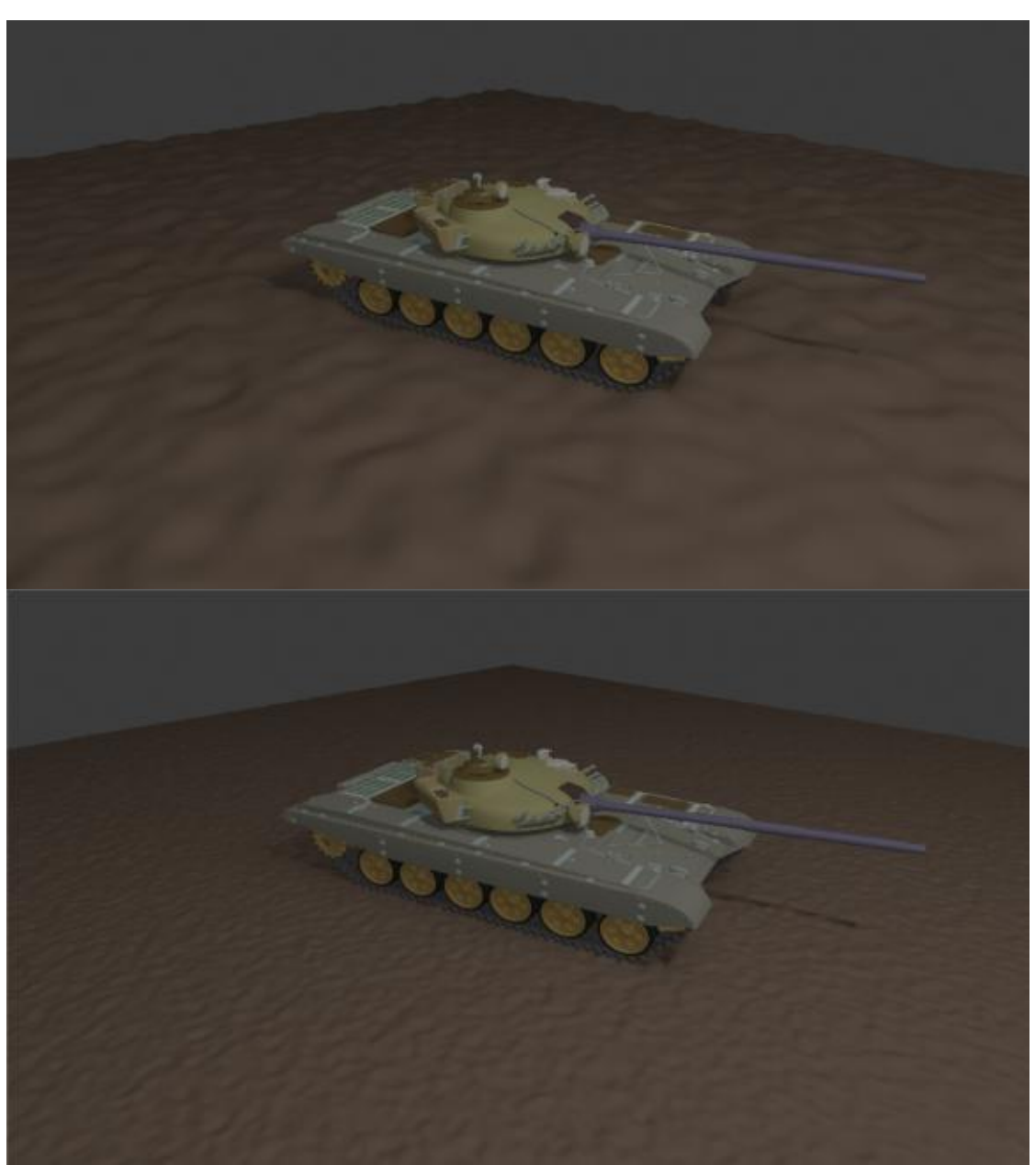

Fig. 5. Visualization of the G7 (top) and G3 (bottom) ground meshes with the T72 target.

## V. Experiments and Results

In this section, we describe our experiments and results. First, in Section A we study how the simulation times of Salsa and its representativity scales with the ray density. Then, in Section B, we study the impact of the EM couplings on the ATR performances. Finally, in Section C we demonstrate that Salsa can generate massive training datasets with a large variety of couplings to make the ATR models more robust.

### *A. Computing Performances and representativity of Salsa*

To study the computing performance and representativity of Salsa, we ran several productions of our 21,600 training images on a single computing node equipped with an Nvidia GeForce RTX 4090 GPU. We consider the G6 ground mesh associated with the M1 materials and the 10 3D models of the targets. For each production, we consider a different ray density in Salsa by varying the surface associated with each ray (the greater the ray surface, the lower the ray density). We measure the simulation time of each production to study the computing performance of Salsa. Then, we train ATR models with each production and test the models on the MSTAR data to study the representativity of the synthetic data.

As shown in Fig. 6, the simulation time of Salsa scales exponentially with the ray surface. With a ray surface of 1e-6 m², the simulation of the whole 21,600 images dataset takes approximately 7 hours, with about one second per image (i.e. close to a real-time acquisition). With a ray surface of 1e-4 m², the whole production time is significantly reduced to about 7 minutes with a throughput of 2e-2 seconds per image. With a ray surface of 1e-3 m² and below, the simulation time reaches its asymptote and takes about 1 minute to produce 21,600 images (approximately 3e-3 sec per image). This asymptote is caused by unavoidable computational overhead for loading the geometry, initializing the ray tracing engine and writing the simulation results.

We observe from Fig. 7 that the ATR accuracy remains remarkably stable for ray surfaces between 1e-6 m² and 1e-4m². The best score is reached at a surface of 1e-4m² with an accuracy of 85 %. For larger ray surfaces like 1e-3m² and 1e-2m², the accuracy drops significantly but remains noticeably high (resp. 81.5 % and 79.5%). However, it collapses at 41.55% for a ray surface of 1e-1m². Although PO removes the need for wavelength-based sampling, overly large ray surfaces lead to insufficient geometric sampling and a loss of 3D model representativity.

These results are consistent with the qualitative analysis of the SAR image obtained (see Fig. 8). For ray surfaces between 1e-6m² and 1e-3m², we observe only minor differences in the target signatures. The signatures start deteriorating significantly for ray surfaces of 1e-2m² and higher.

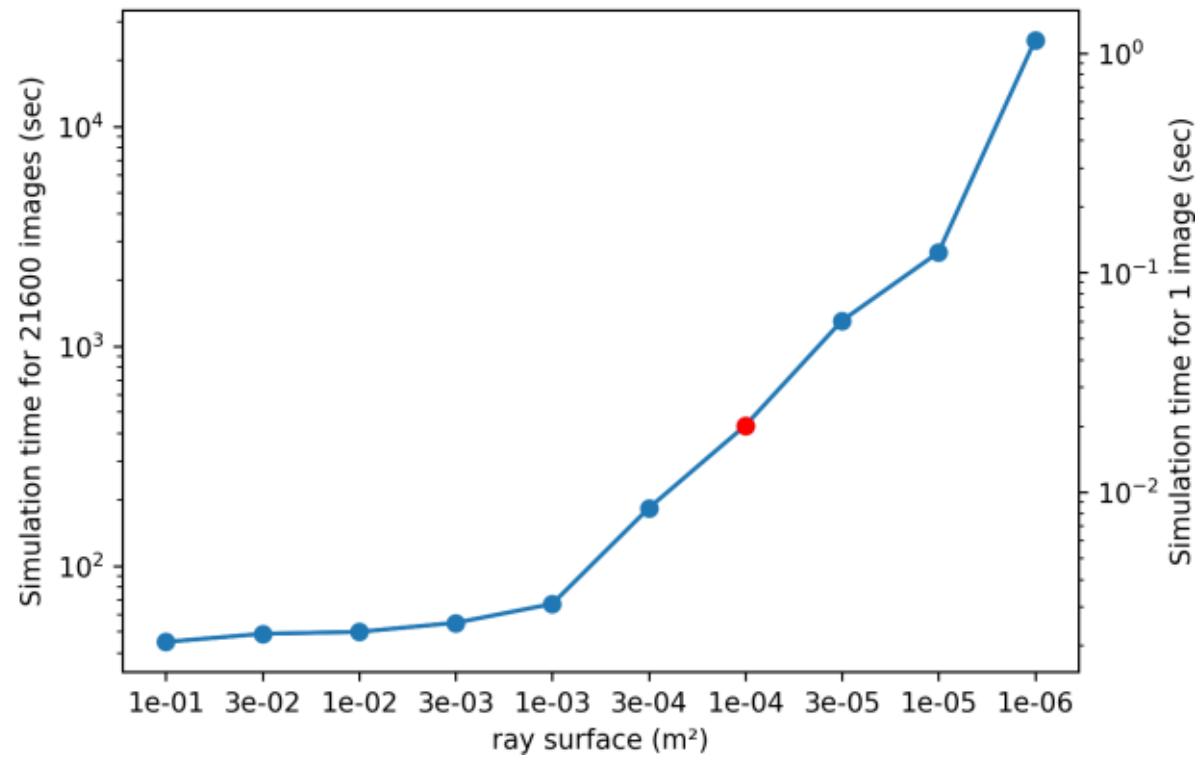

Fig. 6. Salsa simulation time vs ray surface.

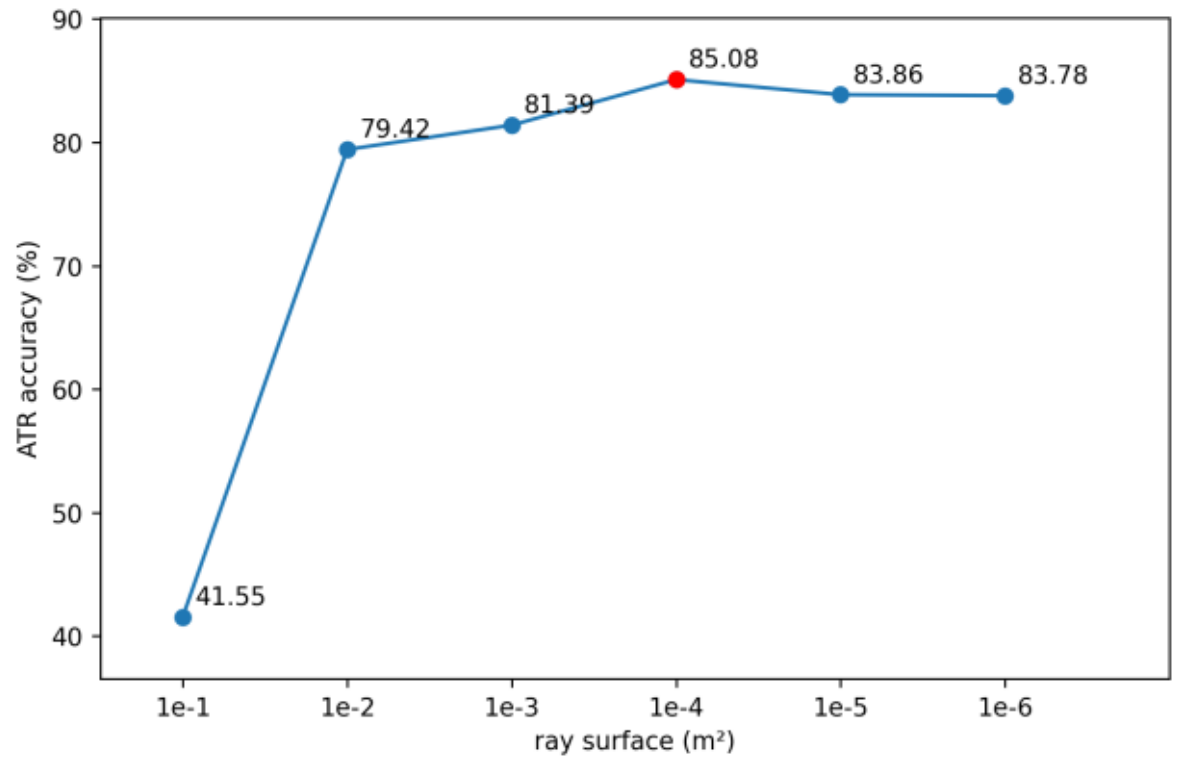

Fig. 7. Accuracy of the ATR models on MSTAR (mean over 5 models trained independantly) vs the ray surface used to produce the synthetic training datasets in Salsa.

To conclude this study, we found that, thanks to its GO/PO physical models, Salsa is robust over several orders of magnitude in ray density. Moreover, we observe that Salsa is very efficient because, even with very high ray densities, the simulation achieves real-time performance. With coarser density, training datasets composed of tens of thousands of images can be generated in a few minutes without loss of representativity. In terms of computing time, the bottleneck becomes the training of ATR models, which takes a few hours per model.

This unlocks new possibilities for ATR, as illustrated in the next sections with the study of the impact of EM coupling on ATR performance. In the following, we use a ray surface of 1e-4 m² as its offers the best trade-of between computing time and simulation representativity (see the red dots on Fig. 6 and Fig. 7).

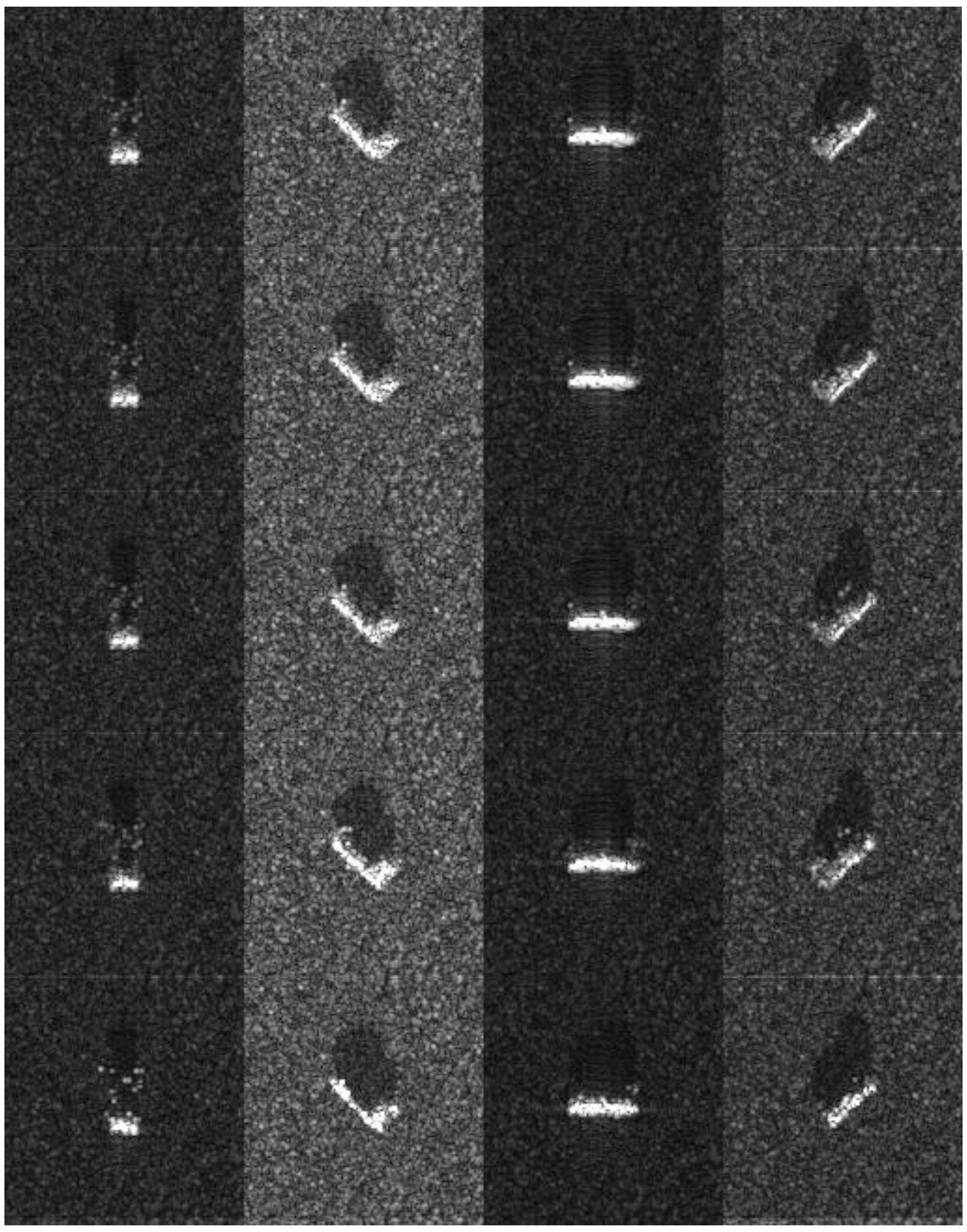

Fig. 8. Comparison on 4 images of the 2S1 target of the simulation results obtained with different ray surfaces (from top to bottom: 1e-6m², 1e-5m², 1e-4m², 1e-3m² and 1e-2m²).

### *B. Impact of EM couplings on ATR performance*

To study the impact of EM couplings on the ATR performance, we ran 10 different productions of our 21,600 images dataset. For each production, we use a different ground mesh. We first test a flat ground for baseline comparison. Then, we successively use the 9 meshes of TABLE I. We considered the M1 EM material of TABLE II. for all grounds. All these productions, which took only 70 minutes to complete, give us 10 datasets of images that differ only in their EM couplings.

With each dataset, we trained 3 ATR models, evaluated them on the MSTAR data, and averaged the resulting accuracies. Thus, by comparing the score obtained with each dataset we can measure the impact of the EM couplings on the ATR performance. We also trained a model on images that don't have EM couplings for baseline comparison.

A qualitative analysis of the images reveals a significant impact of the EM couplings variation on the target signatures (see Fig. 9). As shown in TABLE III. these differences have a significant impact on ATR accuracy. We first observe that the worst results are obtained by ATR models trained on SAR images without EM couplings. This represents a drop of about 8.5 % compared to the optimal value. This indicates that EM couplings must be considered in simulations. The results slightly improve by 1.5 % when considering the EM couplings of a flat ground. However, this is still 7 % below the best results. This shows that it is important to model ground roughness in the simulation. With rough ground surface, the results improve significantly, but we observe important variation of the ATR accuracy (about 4 %) depending on the surface features, although our approach remains robust for most couplings as most accuracies are between 85 % and 86 %. We observed only minor differences across the different EM materials listed in TABLE II. for the ground. This demonstrates that EM couplings have an important impact on ATR results.

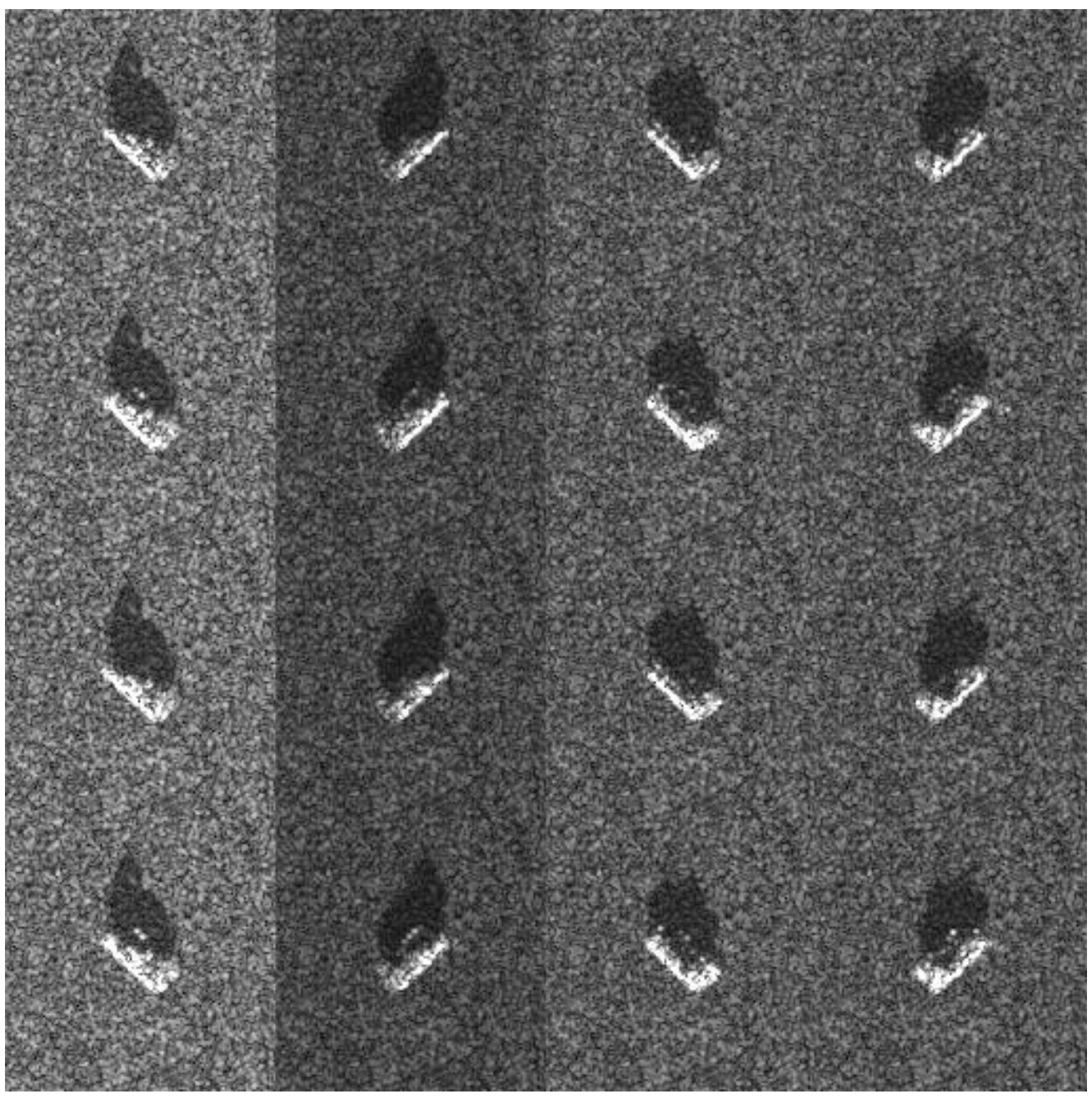

Fig. 9. Comparison on 4 images of the 2S1 target with different ground surface to generate the EM couplings (from top to bottom : no couplings, flat surface, G3 mesh, G4 mesh).

TABLE III. ATR RESULTS ON MSTAR FOR DIFFERENT GROUND SURFACES USED FOR TRAINING.

| Ground ID | ATR score |
|---|---|
| No coupling | 77.93 % |
| Flat ground | 79.43 % |
| G1 | 84.47 % |
| G2 | 85.45 % |
| G3 | 86.42 % |
| G4 | 82.34 % |
| G5 | 83.36 % |
| G6 | 85.08 % |
| G7 | 85.58 % |
| G8 | 85.53 % |
| G9 | 85.93% |

### C. *ATR models robust to EM couplings variation*

It is important to note that, to the author's knowledge, the MSTAR targets were located on the same field, resulting in limited ground diversity. Thus, there might be a single ground mesh that leads to optimal EM couplings for all the MSTAR images. However, in real operational contexts, the ground may vary from one image to another, and we need an ATR model that is robust to EM couplings variation.

To do this, we used all the images generated with our 9 ground meshes to vary the EM coupling during the model training. With this new data-augmentation technique, we randomly choose an EM coupling variant each time an image is selected to be included in a mini-batch. Then, the model is still trained with the same amount of data as previously.

Individual models trained in this way reach an average accuracy of 86.5 %, equaling the score found with optimal EM couplings. We slightly improved this result by averaging the predictions of 5 independent models (i.e. bagging technique), which lead to an accuracy of 87 %.

To show that these models are more robust to the EM coupling variability, we produced several synthetic datasets with different ground couplings. To have different angles than the ones used for training, we adopted the same azimuth and incidence angles as the MSTAR test images. Thus, our test dataset consists of 3,203 images that are the synthetic counterparts of the MSTAR measurements. We produced 9 variants of this dataset by using different ground meshes for each production. These surfaces correspond to new realizations of the statistics described in TABLE I. To simulate the reality gap and make the test target signatures sufficiently different from the training ones, we applied adversarial attacks on the test data [30]. Using these datasets, we evaluated our models trained with the multi-coupling data-augmentation. As shown on TABLE IV. ,we observe small variations of the accuracy with the different coupling, with an average accuracy of 93.59 % across all the datasets.

TABLE IV. ATR RESULTS ON SYNTHETIC TEST DATASETS WITH DIFFERENT GROUND SURFACES FOR MODELS TRAINED WITH MULTI-COUPLING DATA-AUGMENTATION.

| Ground ID used for test | ATR score |
|---|---|
| G1 | 94.59 % |
| G2 | 94.75 % |
| G3 | 94.86 % |
| G4 | 92.38 % |
| G5 | 93.59 % |
| G6 | 93.59 % |
| G7 | 93.71 % |
| G8 | 93.18 % |
| G9 | 91.67 % |

Using the same test datasets, we also evaluated the models of the previous section that have been trained with a single ground surface. As shown on Fig. 10, when these models are evaluated on test data that are based on the same ground statistics as the one used for training, we found an average accuracy of 93.02 % which is close to the 93.59 % average accuracy found for all the test datasets with the multi-coupling data-augmentation. However, when evaluated on test data with different ground statistics than the one used for training, the model's accuracy significantly drops by about 3.67 % on average. In this case, the accuracies are, on average, 4 % below the score obtained with models trained with multi-coupling data-augmentation.

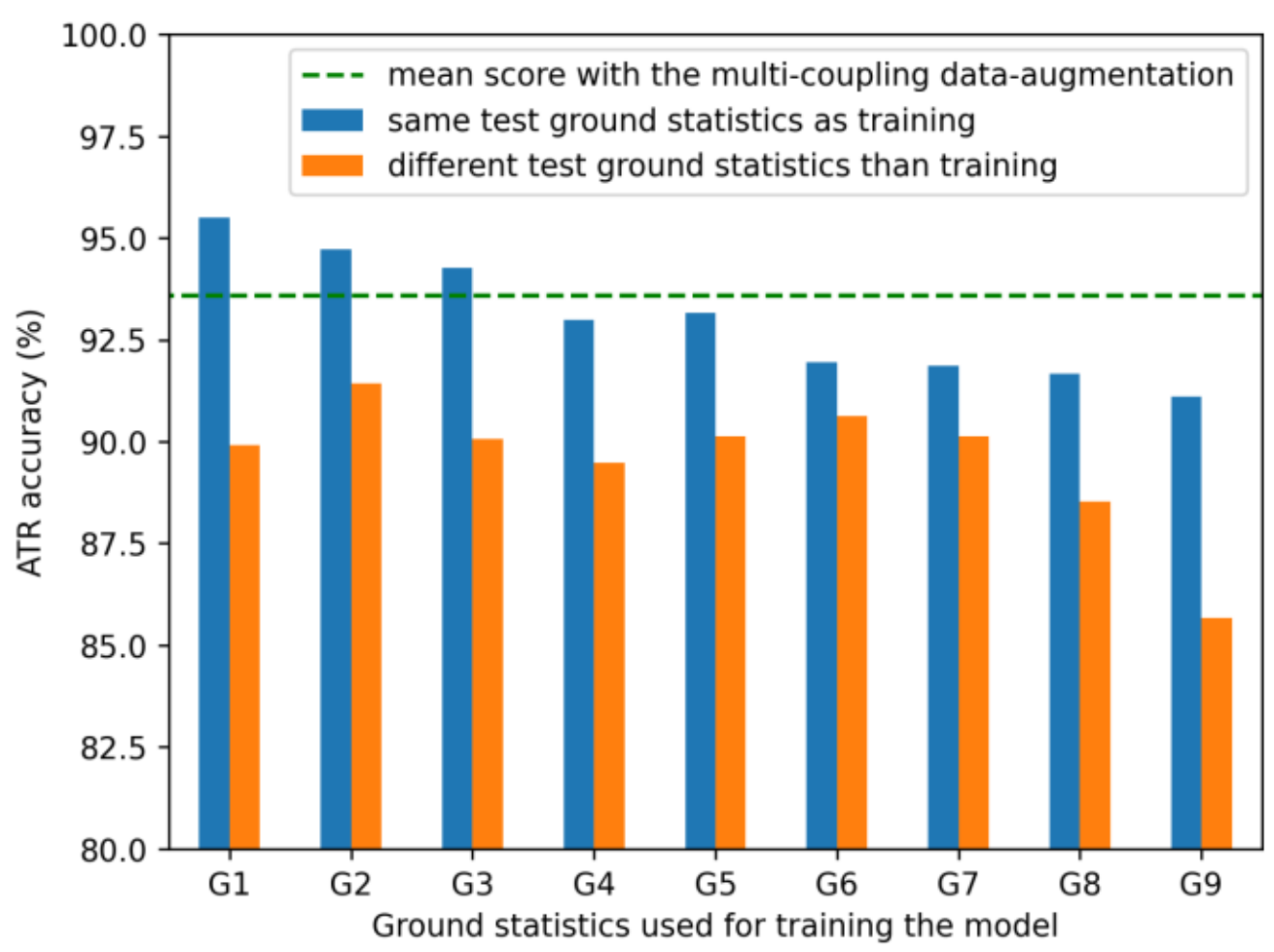

Fig. 10. Comparison of the ATR scores obtained on synthetic test datasets for models trained with single ground surface, and the mean score obtained with the multi-coupling data-augmentation on all the synthetic test datasets.

This demonstrate that introducing EM coupling variability, it is possible to train robust ATR models using synthetic datasets without relying on assumptions regarding a specific ground surface for the targets. By having a fast SAR simulator, we can generate any required variation on demand.

## VI. CONCLUSION

In this work, we have demonstrated that the Salsa SAR simulator can generate representative datasets in a timely manner to train ATR models. Using a single GPU (Nvidia RTX 4090), we were able to generate 21,600 images in 7 minutes. We have shown that models trained with these synthetic data reach an accuracy of 86 % on the MSTAR measurements without fine-tuning the simulations to unrealistically match the ground truth of the test measurements. Thus, starting only with 3D models of objects of interest, we can train ATR models in just a few hours. Generating and labelling the datasets is no longer a bottleneck.

Having such a fast simulator is important because (1) it enables the generation of massive datasets that cover the variability of real measurements to have more robust ATR models, (2) it unlocks new experiments that are crucial for studying ATR, identifying the domain of validity of the models, and improve them.

To illustrate this point, we conducted several experiments to study the sensitivity of ATR models to the variation of EM couplings with generic Gaussian surfaces. In approximately 70 minutes, we produced 10 datasets (of 21,600 images each) using ground meshes with varying roughness to modulate the EM couplings in the images. By training ATR models on these different datasets, we have shown that (1) EM couplings must be considered in the simulations, (2) rough ground surfaces must be considered to have representative EM couplings, (3) the simulated roughness should match the one found in the real measurements, otherwise the test accuracy might drop up to 4 %. As the ground of the targets may vary greatly in a real operational context, we proposed a new data-augmentation technique that uses all our datasets to vary the EM couplings during the training of the models. ATR models trained with this approach reach an accuracy of 87 %, which matches the performance obtained with the best ground roughness. Thus, by covering the EM couplings variability in our training, we obtained ATR models that are more robust.

In future work, we plan to apply this approach to other variation factors. In particular, we will introduce variation in the 3D models of the targets by rotating the articulated parts (e.g. turrets) and adding/removing the removable components (e.g. fuel tanks). We will also consider different EM materials for the target to simulate variation in their EM reflectivity. Furthermore, we will enrich the immediate environment of the target by adding natural or man-made objects (trees, bushes…). To do this, we will integrate Salsa into an existing production pipeline dedicated to massive data-augmentation [31]. We also plan to extend Salsa to consider objects that dynamically move during radar acquisition (e.g. ships) to further enhance representativeness.

## ACKNOWLEDGMENT

This work has been funded by the AID (Agence Innovation Defense) and the DGA (French MoD) in the context of the TACOS project. Thanks to them.

## REFERENCES

[1] Y. Li, et al. "DeepSAR-Net: Deep convolutional neural networks for SAR target recognition," 2017 IEEE ICBDA, 2017, pp. 740-743

[2] D. Morgan. "Deep convolutional neural networks for ATR from SAR imagery." ASARI XXII. Vol. 9475. SPIE, 2015.

[3] J. Quionero-Candela, M. Sugiyama, A. Schwaighofer, and N.D. Lawrence. 2009. Dataset Shift in Machine Learning. The MIT Press.

[4] B. Camus, J. Houssay, C. Le Barbu, E. Monteux, C. Saleun, C. Cochin. Combining SAR Simulators to Train ATR Models with Synthetic Data. 2026. In Proceedings of EUSAR 2026.

[5] Camus, B., Barbu, C. L., & Monteux, E. (2022). Robust SAR ATR on MSTAR with Deep Learning Models trained on Full Synthetic MOCEM data. In Proc. CAID'22.

[6] T. Ross, S. Worrell, V. Velten, et al. , "Standard SAR ATR evaluation experiments using the MSTAR public release data set," Proc. SPIE 3370, ASARI V, 1998

[7] B. Lewis, et al. "A SAR dataset for ATR development: the Synthetic and Measured Paired Labeled Experiment (SAMPLE)." ASARI XXVI. Vol. 10987. SPIE, 2019.

[8] Auer, S.; Hinz, S.; Bamler, R. Ray-tracing simulation techniques for understanding high-resolution SAR images. *IEEE Trans. Geosci. Remot. Sens.* **2010**, *48*, 1445–1456.

[9] Hazlett, M.; Andersh, D.J.; Lee, S.W.; Ling, H.; Yu, C. XPATCH: A high-frequency electromagnetic scattering prediction code using shooting and bouncing rays. In *Targets and Backgrounds: Characterization and Representation*; International Society for Optics and Photonics: Bellingham, WA, USA, 1995; Volume 2469, pp. 266–275.

[10] Andersh, D.; Moore, J.; Kosanovich, S.; Kapp, D.; Bhalla, R.; Kipp, R.; Courtney, T.; Nolan, A.; German, F.; Cook, J.; et al. Xpatch 4: The next generation in high frequency electromagnetic modeling and simulation software. In Proceedings of the Record of the IEEE 2000 International Radar Conference [Cat. No. 00CH37037], Alexandria, VA, USA, 12 May 2000; IEEE: Piscataway, NJ, USA, 2000; pp. 844–849.

[11] Woollard M, Blacknell D, Griffiths H, Ritchie MA. SARCASTIC v2.0—High-Performance SAR Simulation for Next-Generation ATR Systems. Remote Sensing. 2022; 14(11):2561. https://doi.org/10.3390/rs14112561

[12] C. Cochin, P. Pouliguen, B. Delahaye, D. Le Hellard, P. Gosselin, and F. Aubineau "MOCEM - An 'all in one' tool to simulate SAR image," In Proceedings of *EUSAR*, 2008.

[13] C. Cochin, J.-C. Louvigne, R. Fabbri, C. Le Barbu, A. O Knapskog, and N. Ødegaard, "Radar simulation of ship at sea using mocem v4 and comparison to acquisitions," 10 2014.

[14] N. Ødegaard, A. O. Knapskog, C. Cochin and J. Louvigne, "Classification of ships using real and simulated data in a convolutional neural network," 2016 IEEE RadarConf, 2016.

[15] D. Malmgren-Hansen et al., "Improving SAR Automatic Target Recognition Models With Transfer Learning From Simulated Data," in IEEE GRSL, vol. 14, no. 9, Sept. 2017.

[16] M. Cha, et al., "Improving Sar Automatic Target Recognition Using Simulated Images Under Deep Residual Refinements," IEEE ICASSP, 2018.

[17] B. Lewis, J. Liu, A. Wong, "Generative adversarial networks for SAR image realism," Proc. SPIE, Algorithms for Synthetic Aperture Radar Imagery XXV, 1064709, 2018.

[18] B. Camus, E. Monteux and M. Vermet. Refining Simulated SAR images with conditional GAN to train ATR Algorithms. In Proceedings of the Conference on Artificial Intelligence for Defense (CAID), 2020.

[19] N. Inkawhich et al., "Bridging a Gap in SAR-ATR: Training on Fully Synthetic and Testing on Measured Data," in IEEE Journal of Selected Topics in Applied Earth Observations and Remote Sensing, vol. 14, pp. 2942-2955, 2021.

[20] E. Delhommé, H. Remusati, C. Lesueur, J. Petit-Frère Physics-inspired data augmentation for SAR ATR: a new approach to tackle the synthetic-to-measured Domain Gap. In Proc. CAID 2025.

[21] B. Camus, T. Voillemin, C. Le Barbu, J-C. Louvigné, C. Belloni and E. Vallée, "Training Deep Learning Models with Hybrid Datasets for Robust Automatic Target Detection on real SAR images," 2024 International Radar Conference (RADAR), Rennes, France, 2024, pp. 1-6, doi: 10.1109/RADAR58436.2024.10993968.

[22] A. Madry, et al., "Towards deep learning models resistant to adversarial attacks", Proc. ICLR, 2018.

[23] Ian J Goodfellow, Jonathon Shlens, and Christian Szegedy. Explaining and harnessing adversarial examples. arXiv preprint arXiv:1412.6572, 2014

[24] Tobin et al. Domain randomization for transfering deep neural networks from simulation to the real world. IEEE IROS 2017

[25] I. Goodfellow, Y. Bengio and A. Courville, Deep Learning, 2016, [online] Available: http://www.deeplearningbook.org.

[26] Simonyan, K., & Zisserman, A. (2014). Very deep convolutional networks for large-scale image recognition. arXiv preprint arXiv:1409.1556.

[27] Huang, G., Liu, Z., Van Der Maaten, L., & Weinberger, K. Q. (2017). Densely connected convolutional networks. In Proceedings of IEEE CVPR.

[28] Y. Nesterov. A method of solving a convex programming problem with convergence rate o (1/k2). In Sov. Math., Dokl. , vol. 27, pp. 372–376, 1983

[29] Smith, L. N. (2017, March). Cyclical learning rates for training neural networks. In 2017 IEEE WACV

[30] Kurakin, A., Goodfellow, I. J., & Bengio, S. (2018). Adversarial examples in the physical world. In *Artificial intelligence safety and security* (pp. 99-112). Chapman and Hall/CRC.

[31] B. Camus, T. Voillemin, C. Le Barbu, M. Toutirais, J-C. Louvigné, et al.. Génération massive d'images SAR synthétiques pour l'IA à fins de classification automatique (ATR). Journées Environnement Electro-Magnétique (ENVIREM) 2025, Onera, Jun 2025, Palaiseau (91), France.

[32] N. Baghdadi, C. King, A. Chanzy, J-P. Wigneron. An empirical calibration of the integral equation model based on SAR data, soil moisture and surface roughness measurement over bare soils. *International Journal of Remote Sensing*, 2002

[33] F. Ulaby; M. C. Dobson; J. L. Álvarez-Pérez, *Handbook of Radar Scattering Statistics for Terrain* , Artech, 2019.

[34] M. Zribi, N. Baghdadi, N. Holah, O. Fafin, New methodology for soil surface moisture estimation and its application to ENVISAT-ASAR multi-incidence data inversion, Remote Sensing of Environment, Volume 96, Issues 3–4, 2005,